\documentclass[showpacs,amssymb,aps,pre]{revtex4}

\usepackage{epsfig}
\usepackage{graphpap}
\usepackage{rotating}

\newcommand{\ru}[3]{\rule[#1mm]{#2mm}{#3mm}}

\newcommand{\eps}{\varepsilon}

\begin{document}
\setlength{\unitlength}{1mm}

\title{Analytical and Numerical Investigation of the Phase--Locked Loop
with Time Delay}

\author{Michael Schanz}

\affiliation{Institute of Parallel and Distributed Systems (IPVS), University
  of Stuttgart, Breitwiesenstra{\ss}e 20--22, D--70565 Stuttgart, Germany}
\email[E-mail: ]{michael.schanz@informatik.uni-stuttgart.de}

\author{Axel Pelster}

\affiliation{Institute of Theoretical Physics, Free University of
Berlin, Arnimallee 14, D--14195 Berlin, Germany}
\email[E-mail: ]{pelster@physik.fu-berlin.de}

\date{\today}

\begin{abstract}
We derive the normal form for the delay--induced Hopf bifurcation in
the first--order phase--locked loop with time delay by the multiple
scaling method.  The resulting periodic orbit is confirmed by
numerical simulations. Further detailed numerical investigations
demonstrate exemplarily that this system reveals a rich dynamical
behavior.  With phase portraits, Fourier analysis and Lyapunov spectra
it is possible to analyze the scaling properties of the control
parameter in the period--doubling scenario, both qualitatively and
quantitatively.  Within the numerical accuracy there is evidence that
the scaling constant of the time--delayed phase--locked loop coincides
with the Feigenbaum constant $\delta \approx 4.669$ in one--dimensional
discrete systems.
\end{abstract}
\maketitle
\section{Introduction}
Many nonlinear dynamical systems in various scientific disciplines are
influenced by the finite propagation time of signals in feedback
loops. A typical physical system is provided by a laser where the
output light is reflected and fed back to the cavity
\cite{laser1,laser2}. But time delays also occur in biology due to
physiological control mechanisms \cite{neuro1,neuro2} or in economy
where the finite velocity of information processing has to be taken
into account \cite{econo1,econo2}. Furthermore, realistic models in
population dynamics or in ecology include the duration for the
replacement of resources \cite{popu1,popu2}.\\

All these different systems have in common that the inherent time
delay may induce dynamical instabilities. Numerous experimental and
theoretical studies have demonstrated this for the emergence of
oscillatory behavior, quasi-periodicity, chaos or intermittency. But
time--delayed feedbacks can also have the opposite effect. They have
even been devised to stabilize previously unstable stationary states
or limit cycles \cite{stab1,stab2,stab3}. In particular this method
allows to control or to prevent undesirable chaotic behavior in a
time--continuous way. In comparison with the time--discrete method of
Ott, Grebogi, and Yorke \cite{ogi} it can be easier implemented as it
relies on less information of the dynamical system.\\

The choice of a paradigmatic model system for analyzing the
fundamental properties of delay--induced instabilities is determined by
several practical conditions. On the one hand it should be guaranteed
that all observable instabilities are purely a result of time
delays. On the other hand the dynamics should be governed by a simple
model equation and allow for a quantitative comparison between theory
and experiment. These conditions are fulfilled, for instance, by the
Mackey--Glass model \cite{neuro1} which describes quite successfully
the anomalies in the regeneration of white blood cells. However the
underlying nonlinear scalar delay differential equation necessitates
three effective control parameters to account for the experimental
data.\\

In this article we report about analytical and numerical
investigations of the rich dynamical behavior of another model system
which was proposed some time ago by Wischert et al. in
Ref.~\cite{Wolfgang}. It represents the first--order phase--locked loop
(PLL) with time delay which synchronizes the phases of two
oscillators.  In comparison with the Mackey--Glass model, this system
has the advantage that it involves only a single effective control
parameter instead of three. Additionally, it can be realized
electronically under well--defined conditions.  Furthermore, an
extension of the PLL with time delay is capable of describing sensibly
physiological control experiments \cite{Peter}.\\

The experimental set--up for the electronic system of a first--order
phase--locked loop (PLL) is shown in Fig.~3 of Ref.~\cite{Wolfgang}. In many
applications the PLL serves for synchronizing the phases of a
reference oscillator and a voltage--controlled oscillator
(VCO). Thereby the frequency of the output signal of the VCO depends
linearly on the input signal.  The output signals of both oscillators
are multiplied by the aid of a mixer. The induced high--frequency
components are then eliminated by a low--pass filter. The resulting
signal is fed back to the input of the VCO. A delay line between the
VCO and the mixer, implemented analogously or numerically, induces the
time delay $\tau \ge 0$.\\

The dynamical variable of interest is the phase difference $q(t)$
between both incoming signals of the mixer (compare with Fig.~3 of
Ref.~\cite{Wolfgang}). Under quite simple assumptions it becomes
possible to derive a nonlinear scalar delay differential equation for
this phase difference \cite{Wolfgang}:
\begin{equation}
\label{E1}
\frac{d}{d t} q ( t ) = - K \, \sin [ q ( t - \tau ) ] \, .
\end{equation}
The parameter $K \ge 0$ denotes the so-called open loop gain of the
PLL. Performing an appropriate scaling of the time
converts the PLL equation (\ref{E1}) to its standard form
\begin{equation}
\label{E2}
\frac{d}{d t} q ( t ) = -  R \, \sin [ q ( t - 1 ) ] \, ,
\end{equation}
where the two parameters $\tau , K$ are reduced to one effective control
parameter
\begin{eqnarray}
R = K \tau \, .
\end{eqnarray}
Thus varying the delay time $\tau$ corresponds to changing the control
parameter $R$. In this paper we analyze both analytically and
numerically the rich structure of delay--induced instabilities of the
PLL equation (\ref{E2}). In Section \ref{multiple} we derive the
normal form of the Hopf bifurcation by applying the multiple scaling
method. The emerging periodic orbit is confirmed in Section
\ref{verification} by numerical simulations. In Section \ref{results}
we study in detail the period--doubling scenario beyond the Hopf
bifurcation with phase portraits, Fourier analysis and Lyapunov
spectra.

\section{Multiple Scaling Method}\label{multiple}

Applying the synergetic system analysis it was shown in
Ref.~\cite{Wolfgang} that a Hopf bifurcation occurs in the PLL
equation (\ref{E2}) at the critical value $R_c=\pi/2$. In the
following we rederive the normal form of this Hopf bifurcation by
using the multiple scaling method \cite{Michael1}. It represents a
systematic technical procedure to deduce the normal form by using an
ansatz how the respective quantities depend on the smallness parameter
\begin{eqnarray}
\label{small}
\eps = \frac{R-R_c}{R_c} \quad \Longleftrightarrow \quad R = R_c ( 1 + \eps ) \, .
\end{eqnarray}
Although the multiple scaling method has been originally developed for
ordinary or partial differential equations
\cite{ms1,ms2,ms3,Bender,Manneville}, it can be also applied to delay
differential equations (see, for instance, the treatment in
Ref.~\cite{elena}).\\

We start with discussing some properties of the Hopf bifurcation. At
first, we mention that the amplitude of the emerging periodic solution
has a characteristic $\eps^{1/2}$--dependence from the smallness
parameter $\eps$, as can be deduced from the linear stability analysis
already presented in Ref.~\cite{Wolfgang}.  Furthermore, the
trajectory approaches the limit cycle slowly near the instability due
to the phenomenon of critical slowing down. Thus the oscillatory
solution $q(t)$ of the PLL equation (\ref{E2}) is based on two
different time scales. The {\em fast} time scale is provided by the
period $T=2 \pi/\Omega$ of the oscillatory solution, whereas the {\em
slow} one characterizes the amplitude dynamics in the transient
regime. Both time scales are separated by a factor of the smallness
parameter $\eps$, as follows again from the linear stability analysis
\cite{Wolfgang}.  These considerations lead to the following ansatz
for the oscillatory solution after the Hopf bifurcation:
\begin{eqnarray}
q(t) &=& q_{\rm stat} + \eps^{1/2} \left[
q_0^+(t') e^{i \Omega t} + q_0^-(t') e^{-i \Omega t} \right]
+ \eps \left[ q_2^+(t') e^{2 i \Omega t} + q_1(t') +
q_2^-(t') e^{-2 i \Omega t} \right] \nonumber \\ &&
+ \eps^{3/2} \left[
q_4^+(t') e^{3 i \Omega t} + q_3^+(t') e^{i \Omega t}
+ q_3^-(t') e^{-i \Omega t}
+ q_4^-(t') e^{-3 i \Omega t} \right]
+ {\cal O}\left( \eps^2 \right) \, .
\label{ans}
\end{eqnarray}
Here the first time scale $t$ and the second time scale $t'$ are
related via
\begin{eqnarray}
t' = \eps \, t \, ,
\end{eqnarray}
where the smallness parameter $\eps$ denotes the deviation from the
bifurcation point according to (\ref{small}), and the respective
amplitudes have the properties
\begin{eqnarray}
q_k^{\pm}(t') = {q_k^{\mp}{} (t')}^* \, , \quad k = 0,2,3,4 \, ; \quad \quad
q_1(t') = q_1 (t')^* \, .
\end{eqnarray}
Now we insert our ansatz (\ref{ans}) in the PLL equation
(\ref{E2}). By doing so, we have to take into account that the slowly
varying amplitudes $q_i^{\pm}(t')$ have the time derivative
\begin{eqnarray}
\frac{d}{dt}q_i^{\pm}(t') = \eps \, \frac{d}{dt'}q_i^{\pm}(t') 
\end{eqnarray}
and that their time delay results in the expansion
\begin{eqnarray}
q_i^{\pm}(t'-\eps) = q_i^{\pm}(t') - \eps \, \frac{d}{dt'}q_i^{\pm}(t')+ {\cal O} (\eps^2) \, .
\end{eqnarray}
The strategy is then to compare all those terms with each other which
have the same power in the smallness parameter $\eps$.  Thereby we
have to guarantee that in each order the respective Fourier
coefficients compensate each other:
\begin{itemize}
\item In the lowest order $\eps^0$ we have only the frequency $0$
which leads to the equality
\begin{eqnarray}
R_c \, \sin (q_{\rm stat}) =0 \, .
\end{eqnarray}
This fixes the stationary state to be
\begin{eqnarray}
q_{\rm stat} = l \, \pi \, , \quad l = 0 , \pm 1 , \pm 2 , \ldots \,.
\end{eqnarray}
In the following we restrict ourselves to consider the reference state
$q_{\rm stat}^{\rm I}=0$ as the other one $q_{\rm stat}^{\rm II}=\pi$
turns out to be unstable for all values of the control parameter $R$.
\item The order $\eps^{1/2}$ contains only the frequency $\pm \Omega$
with the condition
\begin{eqnarray}
\pm i \Omega \, q_0^{\pm} (t') = - R_c e^ {\mp i \Omega} \, q_0^{\pm} (t') \, .
\end{eqnarray}
As the amplitudes $q_0^{\pm} (t')$ should not vanish, we conclude
\begin{eqnarray}
\label{cheq}
- R_c e^ {\mp i \Omega} \mp i \Omega=0\, .
\end{eqnarray}
This condition coincides with the transcendental characteristic
equation (98) of the linear stability analysis of Ref.~\cite{Wolfgang}
if the eigenvalues $\lambda$ at the instability are identified
according to $\lambda=i \Omega$.  The real part of Eq. (\ref{cheq})
\begin{eqnarray}
\cos \Omega = 0 
\end{eqnarray}
leads to the frequency
\begin{eqnarray}
\label{O}
\Omega = \frac{\pi}{2} \, ,
\end{eqnarray}
whereas the imaginary part results in the critical value of the
control parameter:
\begin{eqnarray}
\label{R}
R_c = \frac{\pi}{2} \, .
\end{eqnarray}
\item The order $\eps$ involves two frequency components. The Fourier
coefficients of the frequency $0$ immediately lead to
\begin{eqnarray}
\label{Q1}
q_1 ( t') = 0 \, ,
\end{eqnarray}
and for the frequency $\pm 2 \Omega$ we obtain
\begin{eqnarray}
\label{g-musc-pll-h1}
\pm 2 i \Omega q_2^{\pm}(t') = - R_c e^{\mp 2 i \Omega} q_2^{\pm}(t') \, ,
\end{eqnarray}
which reduces due to the characteristic equation (\ref{cheq}) to
\begin{eqnarray}
\label{Q2}
q_2^{\pm}(t') = 0 \, .
\end{eqnarray}
\item Also the order $\eps^{3/2}$ consists of two frequency
components. For the frequency $\pm \Omega$ we read off
\begin{eqnarray}
\label{NRR}
\left( 1- R_c e^{\mp i \Omega} \right)  \frac{d}{dt'}q_0^{\pm}(t')
= \left(- R_c e^{\mp i \Omega} \mp i \Omega \right) q_3^{\pm}(t') -
\frac{1}{2} R_c e^{\mp i \Omega}
\left[ 2 q_0^{\pm}(t') - {q_0^{\pm}(t')}^2 q_0^{\mp}(t') \right] \, .
\end{eqnarray}
The factor in front of $q_3^{\pm}(t')$ vanishes because of the
characteristic equation (\ref{cheq}). Thus the functions
$q_3^{\pm}(t')$ are not determined in this order, they only follow
from the next order $\eps^2$ and the frequency $\pm \Omega$.  Taking
into account the characteristic equation (\ref{cheq}), we yield from
(\ref{NRR}) the normal form of the order parameter equation:
\begin{eqnarray}
\label{NF2a}
\frac{d}{dt'}q_0^{\pm}(t') = A^{\pm} q_0^{\pm}(t') + B^{\pm}
{q_0^{\pm}(t')}^2 q_0^{\mp}(t') \, .
\end{eqnarray}
There the parameters $A^{\pm}$ and $B^{\pm}$ are defined by
\begin{eqnarray}
\label{NF2b}
A^{\pm} = \frac{i R_c}{1+i R_c} \, , \quad B^{\pm} = - \frac{A^{\pm}}{2} \, .
\end{eqnarray}
Note that the normal form (\ref{NF2a}) of the multiple scaling method
and the normal form of the synergetic system analysis in
Ref.~\cite{Wolfgang} do not coincide, however, they can be mapped into
each other using an appropriate coordinate transformation
\cite{Michael1}.  Correspondingly, we obtain for the frequency $\pm 3
\Omega$
\begin{eqnarray}
\pm 3 i \Omega q_4^{\pm}(t') = \frac{1}{6} R_c e^{\mp 3 i \Omega}
\left[ {q_0^{\pm}(t')}^3 - 6 q_4^{\pm}(t') \right] \, ,
\end{eqnarray}
which reduces due to (\ref{cheq}), (\ref{O}) and (\ref{R}) to
\begin{eqnarray}
q_4^{\pm}(t') = \frac{1}{24} {q_0^{\pm} (t')}^3 \, .
\end{eqnarray}
As the functions $q_3^{\pm}(t')$ and $q_4^{\pm}(t')$ are of the order
$\eps^{3/2}$ according to the ansatz (\ref{ans}), they are irrelevant
for the order $\eps$ in which we are interested.
\end{itemize}
It remains to solve the order parameter equation (\ref{NF2a}),
(\ref{NF2b}) by using polar coordinates
\begin{eqnarray}
q_0^{\pm} (t') = R (t') \, e^{\pm i \varphi ( t')} \, .
\end{eqnarray}
The resulting stationary solution turns out to be
\begin{eqnarray}
\label{SOL}
R (t')= \sqrt{2} + {\cal O} (\eps^2)
\end{eqnarray}
together with the phase
\begin{eqnarray}
\label{phase}
\varphi(t) = \Omega (\eps)t + \varphi_0  \, .
\end{eqnarray}
Here the frequency turns out to be
\begin{eqnarray}
\label{oom}
\Omega (\eps) = R_c + {\cal O} (\eps^2) \, .
\end{eqnarray}
Thus we conclude from (\ref{O}), (\ref{Q1}), (\ref{Q2}), (\ref{SOL}), (\ref{phase}), and (\ref{oom})
that the oscillatory solution after the Hopf bifurcation reads 
\begin{eqnarray}
\label{q}
q(t) = c_0(\eps) + c_1(\eps) \cos \left[ \varphi(t) + \psi_1 \right] + c_2(\eps) \cos \left[ 2 \varphi(t) + \psi_2 \right] +
{\cal O} \left( \eps^{3/2} \right) \, ,
\end{eqnarray}
where the respective coefficients read
\begin{eqnarray}
\label{coeff}
c_0 (\eps) = 0 + {\cal O} \left( \eps^2 \right) \, , \quad
c_1 (\eps) = \sqrt{8 \eps} + {\cal O} \left( \eps^{3/2} \right) \, , \quad
c_2 (\eps) = 0 + {\cal O} \left( \eps^2 \right) \, .
\end{eqnarray}
It coincides with the result of the synergetic system analysis in
Ref.~\cite{Wolfgang} up to the order $\eps$. Although we restrict
ourselves to this order, the systematics of the multiple scaling
method is obvious, thus an extension of the ansatz (\ref{ans}) to
higher orders is straight--forward, but the calculation would
become quite cumbersome.
\section{Numerical Verification}\label{verification}

In order to numerically verify our analytical result, we integrated
the underlying PLL equation (\ref{E2}). By doing so, we varied the
control parameter $R$ in the vicinity of the instability $R_c=\pi/2$
in such a way that the smallness parameter $\eps = (R-R_c)/R_c$ took
200 equidistant values between $10^{-5}$ and $10^{-1}$. We used a
Runge--Kutta--Verner method of the IMSL library as an integration
routine and performed a linear interpolation between the respective
values. In particular in the immediate vicinity of the instability the
phenomenon of critical slowing down led to a transient behavior. To
exclude this, we iterated the discretized delay differential equation
for each value of the control parameter at least $10^6$
times. Afterwards we calculated the power spectrum with a complex FFT
so that the basic frequency $\Omega$ of the oscillatory solution could
be determined with high resolution. Then we performed a real FFT with
the period $T=2 \pi / \Omega$ of the simulated periodic signal
$q(t)=q(t+T)$:
\begin{eqnarray}
\label{F1}
q(t) = \frac{a_0}{2} + \sum_{k=1}^{\infty} \left[ a_k \cos \left( k \Omega t \right) + b_k \sin \left( k \Omega t \right) \right] \, .
\end{eqnarray}

\begin{table}
\begin{center}
\begin{tabular}{|c||c|c||c|c||c|c||}\hline
\begin{minipage}{2cm}
\begin{center}
\rule[0mm]{0mm}{3mm}investigated\\ quantity\\
\end{center}
\end{minipage} &
\multicolumn{2}{|c||}{
\begin{minipage}{2cm}
\begin{center}
analytical \\ expression
\end{center}
\end{minipage}} &
\multicolumn{2}{|c||}{
\begin{minipage}{3cm}
\begin{center}
analytical \\ value
\end{center}
\end{minipage}} &
\multicolumn{2}{|c||}{
\begin{minipage}{3cm}
\begin{center}
numerical \\ value
\end{center}
\end{minipage}} \\
\rule[-2mm]{0mm}{1mm}$A$ & \hspace*{1mm} $A(0)$ \hspace*{1mm} & \hspace*{1mm} $\left. \frac{dA}{d \eps} \right|_{\eps=0}$ \hspace*{1mm} 
& \hspace*{1mm} $A(0)$ \hspace*{1mm} & \hspace*{1mm} $\left. \frac{dA}{d \eps} \right|_{\eps=0}$ \hspace*{1mm}  
& \hspace*{1mm} $A(0)$ \hspace*{1mm} & \hspace*{1mm} $\left. \frac{dA}{d \eps} \right|_{\eps=0}$ \hspace*{1mm}  \\ \hline \hline
\ru{-3}{0}{9}$\Omega$ & ${\displaystyle \frac{\pi}{2}}$ & $0$ &
$1.5708$ & $0.0$ & $1.5708$ & $10^{-8}$ \\ \hline
\ru{-3}{0}{9}$c_0$ & $0$ & $0$ & $0.0$ & $0.0$ &
$-3 \cdot 10^{-4}$ & $-2 \cdot 10^{-3}$ \\ \hline
\ru{-3}{0}{9}$\ln c_1$ & $\frac{1}{2}\ln 8$ & ${\displaystyle \frac{1}{2}}$ &
$1.0397$ & $0.5$ & $1.0356$ & $0.4999$ \\ \hline
\ru{-3}{0}{9}$c_2$ & $0$ & $0$ &
$0.0$ & $0.0$ & $2 \cdot 10^{-4}$ & $6 \cdot 10^{-2}$ \\ \hline
\end{tabular}
\end{center}
\caption{\label{values} Comparing the analytical and the
numerical values for the frequency $\Omega(\eps)$ and the Fourier
coefficients $c_0(\eps)$, $c_1(\eps)$, $c_2(\eps)$ of the oscillatory solution of the
PLL equation after the Hopf bifurcation.}
\end{table}
The Fourier coefficients follow from integrations with respect to one
period $T=2 \pi / \Omega$:
\begin{eqnarray}
a_k =  \frac{2}{T} \int\limits_0^{T} d t \, f ( t ) \,  \cos \left( k \Omega t \right) \, , \quad
k = 0 , 1 , \ldots , \infty \, ; \quad 
b_k  =  \frac{2}{T} \int\limits_0^{T} d t \, f ( t ) \,  \sin \left( k \Omega t \right) \, , \quad
k = 1 , \ldots , \infty \, . 
\end{eqnarray}
From (\ref{F1}) follows then the spectral representation
\begin{eqnarray}
\label{specrep}
q ( t ) = c_0 + \sum_{k=1}^{\infty} c_k \cos \left( k \Omega t +
\phi_k \right)
\end{eqnarray}
with the quantities
\begin{eqnarray}
c_0 = \frac{a_0}{2} \, , \quad c_k = \sqrt{a_k^2 + b_k^2} \, , \quad
\phi_k = - \mbox{arctan}\, \frac{b_k}{a_k} \, , \quad k = 1 , \ldots ,
\infty \, .
\end{eqnarray}
Thus our analytical result (\ref{phase}), (\ref{q}) can be interpreted
as the first terms within a spectral representation (\ref{specrep}),
where the frequency $\Omega = 2 \pi / T$ and the Fourier coefficients
$c_0$, $c_1$, $c_2$ are given by (\ref{oom}) and
(\ref{coeff}). Analyzing the Hopf bifurcation with a FFT, we
numerically determinded $\Omega$, $c_0$, $c_1$, $c_2$ as a function of
the smallness parameter $\eps$.  Comparing the respective numerical
and analytical results, we observe some deviations for small and for
large values of the smallness parameters $\eps$. The former are due to
the phenomenon of critical slowing down, i.e. the system stays longer
in the transient state when the instability is approached, and the
latter arise from the neglected higher order corrections in the
analytical approach.  Therefore we restricted our numerical analysis
to the intermediate interval $[10^{-5},10^{-1}]$ of the smallness
parameter $\eps$.\\

In Tab.~\ref{values} we see that the analytical and numerical
determined quantities agree quantitatively very well. Thus our weakly
non-linear analysis for the delay--induced Hopf bifurcation in the PLL
equation is numerically verified up to $\eps \approx 10^{-1}$.
However, this successfully tests only the order parameter concept for delay
systems, as the lowest nonlinear term in the scalar delay differential
equation of the PLL (\ref{E2}) is a cubic one. Therefore we analyzed
the Wright equation \cite{Wright} with a quadratic nonlinearity in a
separate publication \cite{Michael2}, where we could successfully test
not only the order parameter concept but also the slaving principle,
i.e. the influence of the center manifold on the order parameter
equations. Thus we demonstrated with the Wright equation the validity
of the circular causality chain of synergetics for the Hopf
bifurcation of a delay differential equation.

\section{Further Numerical Results}\label{results}

In the following we summarize various simulations which have been
performed for the delay differential equation (\ref{E2}) of a PLL with
some initial function $q ( t )$ for $-1 \le t \le 0$
\cite{Michael1,Michael3}.  In order to check the quality of the
numerical results, standard integration routines of the Runge--Kutta
type have been applied with different discretizations by adequately
taking into account the delay effects.\\

\begin{figure}[ht]
\centerline{\epsfig{figure=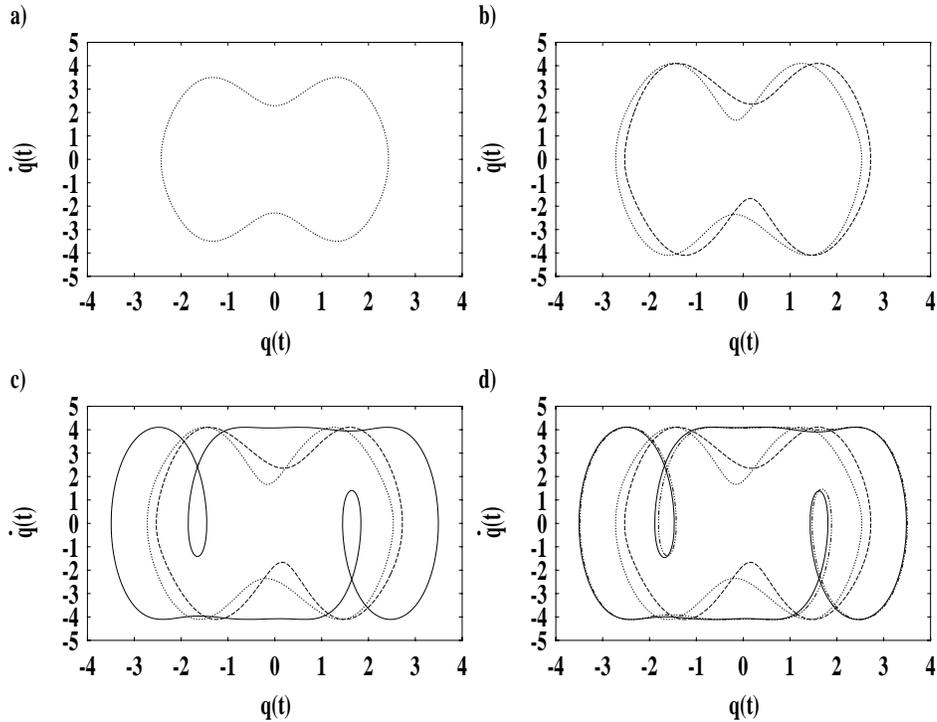,width=10cm,angle=-90}}
\caption{\label{phasep} Phase portraits depicting several limit cycles:
a) $R=3.5$, b) $R=4.1$, c) $R=4.105$ d) $R=4.11$.}
\end{figure}

Firstly, it turns out that the trajectory $q ( t )$ is restricted for
all times to the interval $[- \pi, + \pi]$ if the effective control
parameter $R$ is increased from 0 to about 4.9. If the transient
behavior has decayed, the resulting asymptotic dynamics could be
classified as follows. For $0 \le R \le \pi/2$ there exist two
stationary states, a stable one $q_1 = 0$ and an unstable one $q_2 =
\pi$. At $R = \pi/2$ a super--stable Hopf bifurcation occurs where the
previously stable stationary state $q_{\rm stat}^{\rm I} = 0$ becomes
unstable and a new stable limit cycle emerges \cite{Wolfgang}. In the
range $\pi/2 \le R \lessapprox 3.77$ this oscillatory solution shows a
conspicuous point symmetry with respect to the origin of the phase
portrait in Fig.~\ref{phasep}a. This symmetry is broken at $R \approx
3.77$ as the limit cycle splits into two coexisting limit cycles
\cite{Chrisi}. They are depicted in Fig.~\ref{phasep}b as the
asymptotic dynamics of the initial functions $q ( t ) = \pm 1$ for $-1
\le t \le 0$, respectively. Both coexisting limit cycles are symmetric
to each other concerning the point symmetry with respect to the origin
and remain stable up to $R \approx 4.9$. Note that the instability at
$R \approx 3.77$ was not detected during the initial investigations in
Ref.~\cite{Wolfgang} as there only Fourier spectra were analyzed.\\

\begin{figure}[ht]
\begin{center}
\begin{picture}(120,85)
%\graphpaper[5](0,0)(120,85)
\put(0,85){\epsfig{figure=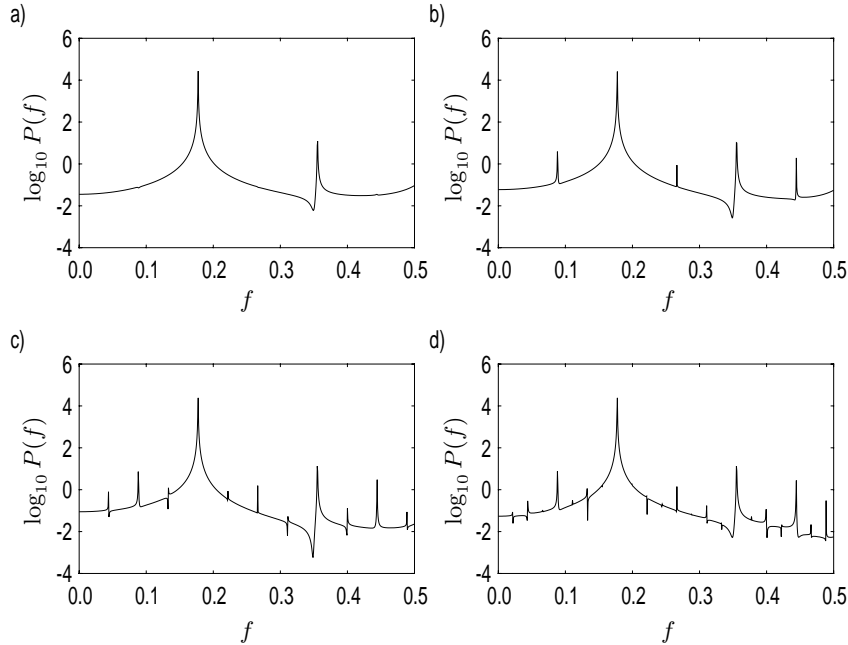,angle=-90,width=12.6cm}}
\put(8,15){\begin{rotate}{90}{$\log_{10} P(f)$}\end{rotate}}
\put(8,58){\begin{rotate}{90}{$\log_{10} P(f)$}\end{rotate}}
\put(64,15){\begin{rotate}{90}{$\log_{10} P(f)$}\end{rotate}}
\put(64,58){\begin{rotate}{90}{$\log_{10} P(f)$}\end{rotate}}
\put(34.5,0){$f$}
\put(34.5,44){$f$}
\put(90.5,0){$f$}
\put(90.5,44){$f$}
\end{picture}
\end{center}
\caption{\label{fourier} Power spectra indicating period--doublings:
a) $R=4.157$, b) $R=4.165$, c) $R=4.1725$, d) $R=4.17375$.}
\end{figure}

\begin{figure}[t]
\begin{center}
\begin{picture}(95,70)
%\graphpaper[5](0,0)(95,70)
\put(0,73.5){\epsfig{figure=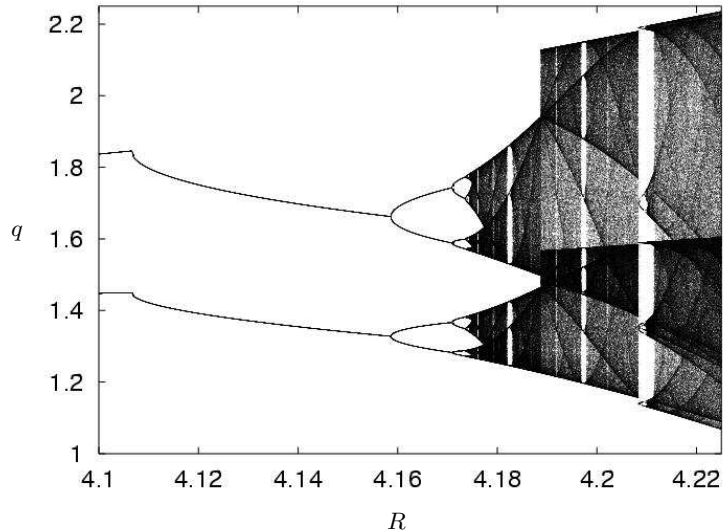,angle=-90,width=10cm}}
\put(0,39){$q$}
\put(50,0){$R$}
\end{picture}
\end{center}
\caption{\label{poin} Bifurcation diagram obtained by a Poincar\'e
section which is defined by the conditions $\dot{q}(t) = 0$ and $q(t)
\in [1,2]$.}
\end{figure}

Increasing the effective control parameter leads to a further
bifurcation at $R \approx 4.105$. Fig.~\ref{phasep}c illustrates that
a new limit cycle emerges with the initial function $q ( t ) = 2$ for
$-1 \le t \le 0$ which coexists for $4.105 \lessapprox R \lessapprox
4.11$ with the two other oscillating solutions. Also this new limit
cycle exhibits a point symmetry with respect to the origin of the
phase portrait. At $R \approx 4.11$ this limit cycle splits into two
new oscillating solutions with the initial functions $q ( t ) = \pm 2$
for $-1 \le t \le 0$, so that the point symmetry is again destroyed
(compare Fig.~\ref{phasep}d).  It turns out that both of them pass
through a separate period--doubling scenario for $4.11 \lessapprox R
\lessapprox 4.175$. This is shown qualitatively by the Power spectra
(see Fig.~\ref{fourier}) for the first three period--doublings. Each
of these bifurcations leads to a subsequent sub--harmonic and to
corresponding higher combination frequencies. The bifurcation diagram
in Fig.~\ref{poin} is an overview over this period--doubling
scenario. It was obtained by Poincar\'e sections of the trajectories
using the software package {\bf AnT~4.669} \cite{AnTpaper,AnTWWW},
whereby the Poincar\'e conditions were $\dot{q}(t) = 0$ and $q(t) \in
[1,2]$.

In order to analyze a period--doubling scenario more quantitatively,
it is advantageous to determine the Lyapunov exponents of the
underlying dynamics. In our case this necessitates to use the common
concept of Lyapunov exponents \cite{wolf} and to extend it to delay
differential equations \cite{Michael1,farmer}. Thereby we have to take
into account that their numerical integration is based on a
discretization procedure. As a consequence, the determination of
Lyapunov exponents for time--delayed dynamical systems is reducible to
the calculation of Lyapunov exponents for a high--dimensional
time--discrete mapping. \\

\begin{figure}[t]
\begin{center}
\begin{picture}(170,60)
%\graphpaper[5](0,0)(170,60)
\put(5,63){\epsfig{figure=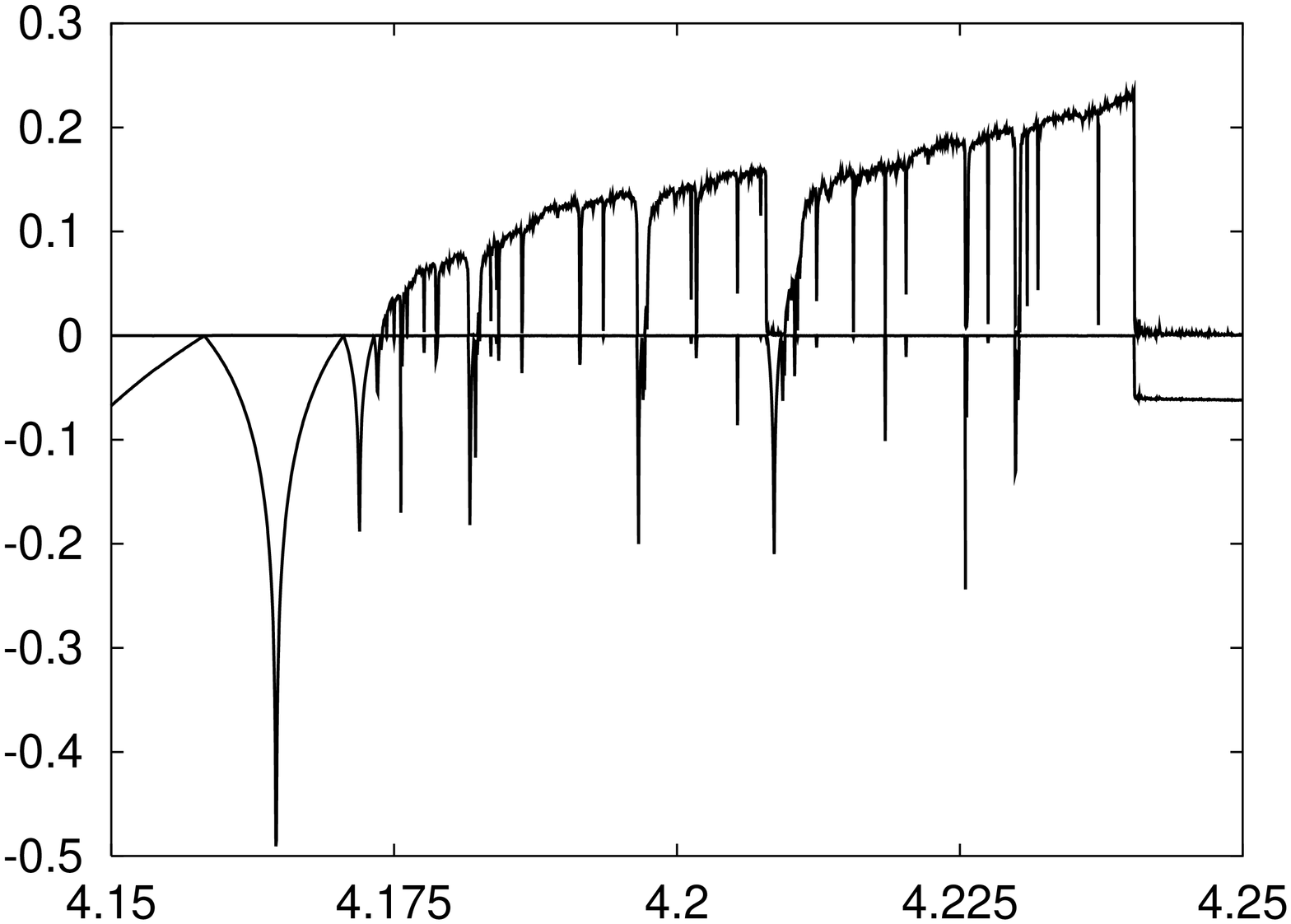,angle=-90,width=85mm}}
\put(50,1){$R$}
\put(7,30){\begin{rotate}{90}{\boldmath $\lambda_1,\lambda_2$}\end{rotate}}
\put(90,63){\epsfig{figure=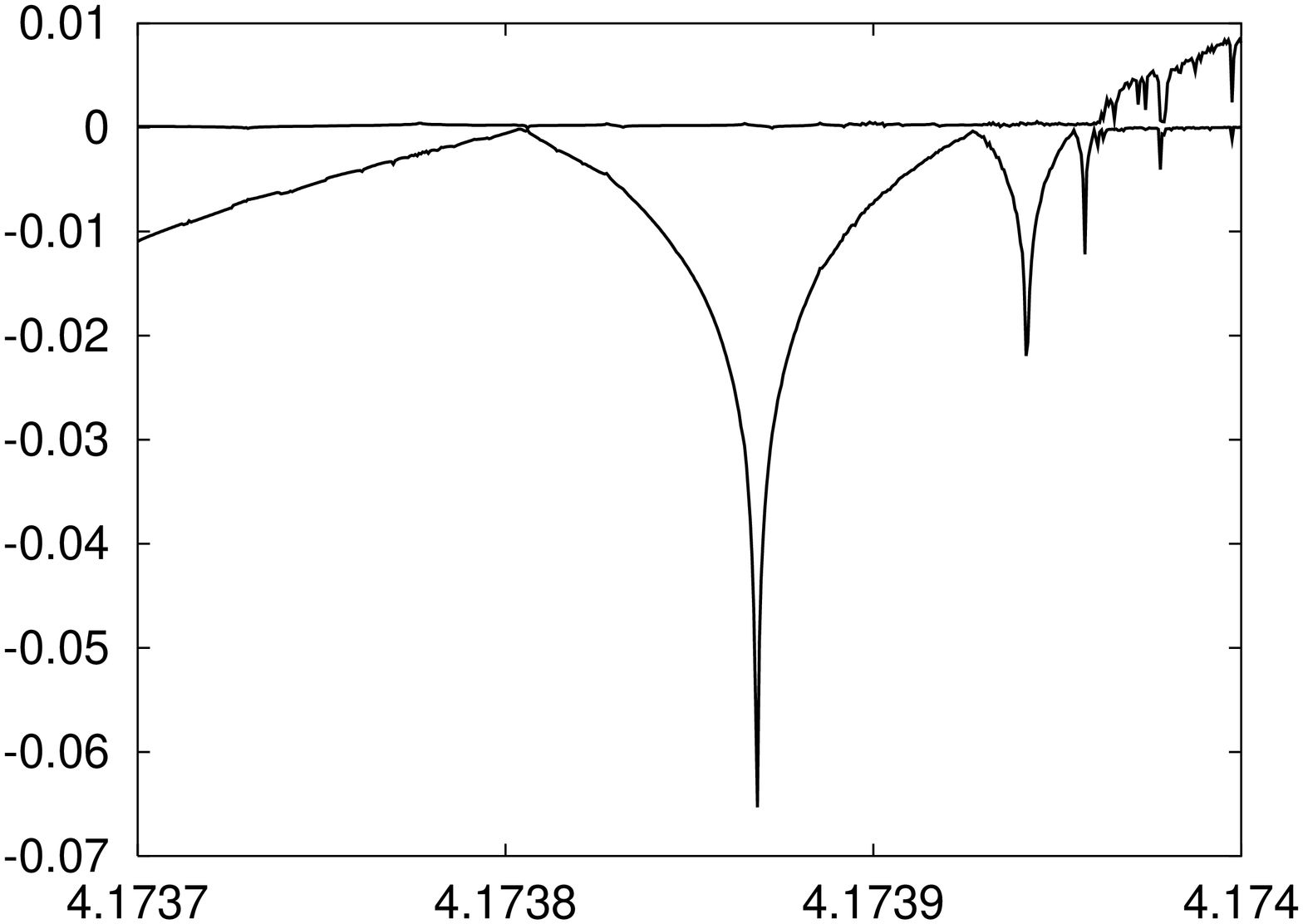,angle=-90,width=85mm}}
\put(135,1){$R$}
\put(92,30){\begin{rotate}{90}{\boldmath $\lambda_1,\lambda_2$}\end{rotate}}
\end{picture}
\end{center}
\caption{\label{lyapunov} Lyapunov spectra illustrating
self--similarity. Shown are the two largest Lyapunov exponents $\lambda_1$
and $\lambda_2$.}
\end{figure}

Fig.~\ref{lyapunov}a shows the two largest Lyapunov exponents within
and above the period--doubling scenario of the PLL. The enlargement of
Fig.~\ref{lyapunov}b clearly reveals the self--similarity of the
spikes and the characteristic scaling properties. One of both Lyapunov
exponents always vanishes due to the moving reference frame. The zeros
of the second Lyapunov exponent coincide with the critical values of
the effective control parameters $R_0, R_1, R_2 , \ldots$ where a
period--doubling occurs.  Table~\ref{feigenbaumzahl} lists the first
bifurcation points and the scaling constants
\begin{equation}
\delta_n = \frac{R_{n - 1} - R_n}{R_n - R_{n + 1}}
\end{equation}
of the effective control parameter. Within the numerical accuracy
there is evidence that the scaling constants $\delta_n$ converge to
the Feigenbaum constant $\delta \approx 4.669$ as in the case of
one--dimensional time--discrete systems
\cite{grossmann,feigenbaum}. If we assume this to be true then we can
estimate the end of the period--doubling scenario
\begin{equation}
R_{\infty} = \frac{\delta R_{n + 1} - R_n}{\delta - 1}
\end{equation}
from the bifurcation points in Tab.~\ref{feigenbaumzahl}. The finding
$R_{\infty} \approx 4.173961$ agrees quite well with the enlarged
Lyapunov spectrum in Fig.~\ref{lyapunov}b.\\

\begin{table}
\begin{center}
\begin{tabular}{|c|c|c|c|}
\hline
$n$\rule[-5mm]{0mm}{12mm} & parameter $R_n$ &
\begin{minipage}{32mm}
\begin{center}period before \\ bifurcation \end{center}
\end{minipage} &
$\delta_n$ \\
\hline
$0$ & $4.158 \pm 5 \cdot 10^{-5}$ & $1$ & $-$ \\
\hline
$1$ & $4.1705 \pm 5 \cdot 10^{-5}$ & $2$ & $4.63
\pm 2 \cdot 10^{-1}$ \\
\hline
$2$ & $4.1732 \pm 5 \cdot 10^{-5}$ & $4$ & $4.49
\pm 5 \cdot 10^{-1}$ \\
\hline
$3$ & $4.173802 \pm 5 \cdot 10^{-7}$ & $8$ & $4.81
\pm 4 \cdot 10^{-1}$ \\
\hline
$4$ & $4.1739272 \pm 2 \cdot 10^{-7}$ & $16$ & $4.76
\pm 10^{-1}$ \\
\hline
$5$ & $4.1739535 \pm 2.5 \cdot 10^{-7}$ & $32$ & $4.46
\pm 3 \cdot 10^{-1}$ \\
\hline
$6$ & $4.1739594 \pm 5 \cdot 10^{-8}$ & $64$ & \hspace*{1mm}$4.72
\pm 4 \cdot 10^{-1}$\hspace*{1mm} \\
\hline
 \hspace*{1mm}$7 \hspace*{1mm}$ & \hspace*{1mm} $4.17396065 \pm 2.5 \cdot 10^{-8}$\hspace*{1mm} & $128$ & $-$ \\
\hline
\end{tabular}
\vspace*{0.3cm}
\end{center}
\caption{\label{feigenbaumzahl} Bifurcation points of the
period--doubling scenario.}
\end{table}

\begin{figure}[ht]
\begin{center}
\begin{picture}(130,95)
%\graphpaper[5](0,0)(130,95)
\put(0,95){\epsfig{figure=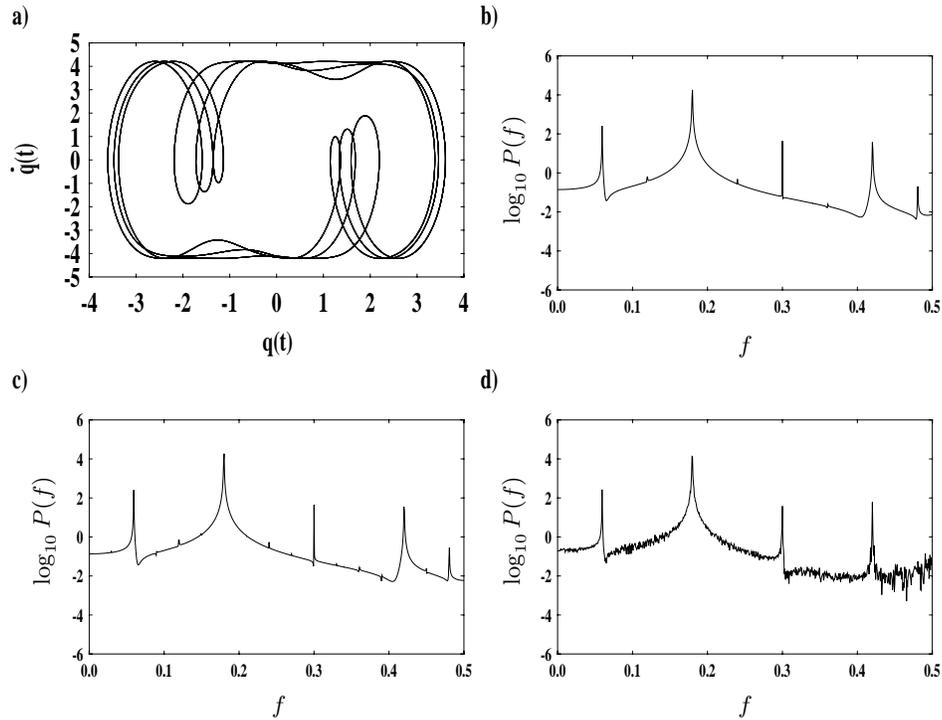,width=10cm,angle=-90}}
\put(9.5,17){\begin{rotate}{90}{$\log_{10} P(f)$}\end{rotate}}
\put(72,17){\begin{rotate}{90}{$\log_{10} P(f)$}\end{rotate}}
\put(72,65){\begin{rotate}{90}{$\log_{10} P(f)$}\end{rotate}}
\put(101,48){$f$}
\put(39,0){$f$}
\put(101,0){$f$}
\end{picture}
\end{center}
\caption{\label{pw} Analyzing the periodic window at $4.2095
\lessapprox R \lessapprox 4.215$: a,b) $R=4.211$, c) $R=4.2115$, d) $R=4.213$.}
\end{figure}

As typical for the period doubling route to chaos, there exist not
only the period doubling scenario below the critical value, that is
for $R < R_{\infty}$, which ends at the critical value $R_{\infty}$
with the emergence of a Feigenbaum attractor, but also the
band--merging scenario with periodic windows above the critical value,
that is for $R > R_{\infty}$. As an example the periodic behavior in
the window $4.2095 \lessapprox R \lessapprox 4.215$ was
analyzed more carefully. The phase portrait of Fig.~\ref{pw}a and the
Power spectrum of Fig.~\ref{pw}b show that at $R \approx 4.2095$ a
limit cycle of period 3 emerges. At $R \approx 4.2115$ it starts to
pass through a period--doubling scenario (see Fig.~\ref{pw}c). Finally,
at $R \approx 4.213$ a chaotic attractor emerges whose power spectrum
in Fig.~\ref{pw}d clearly reveals the structure of the limit cycle of
period 3. This chaotic regime ends at $R \approx 4.2405$ when a global
bifurcation or a transition to transient chaos occurs. For $4.2405
\lessapprox R \lessapprox 4.85$ phase portraits and Power spectra
show that only those limit cycles coexist which emerged at $R \approx
3.77$. At $R \approx 4.85$ two new limit cycles of period 2 are
generated which coexist for $4.85 \lessapprox R \lessapprox 4.90$.\\

\begin{figure}[ht]
\centerline{\epsfig{figure=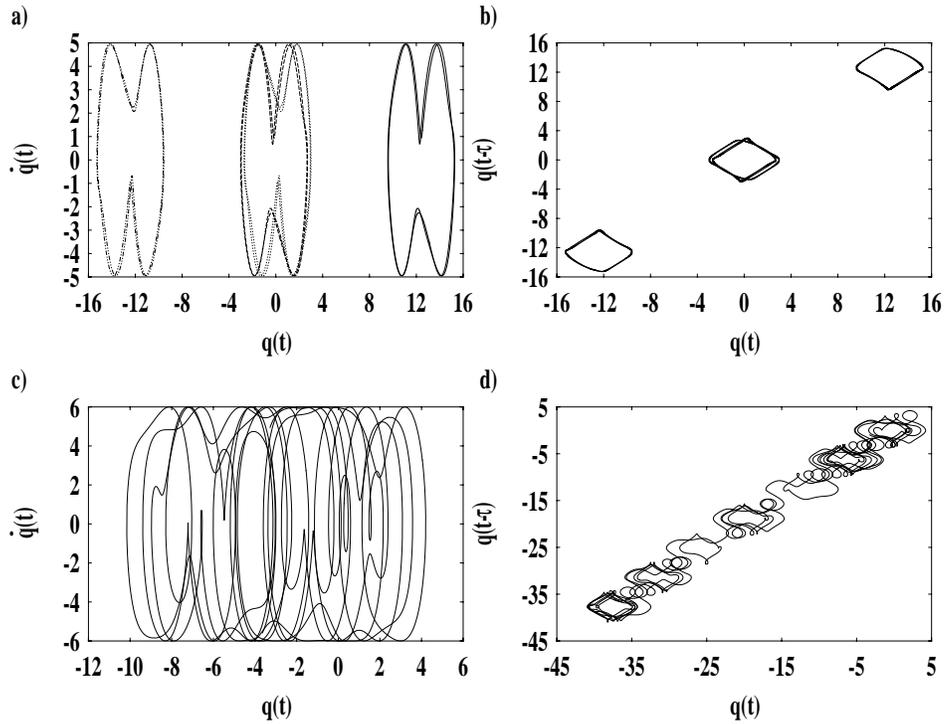,width=10cm,angle=-90}}
\caption{\label{cb} Analyzing the regime $R > 4.90$: a,b) $R=4.95$, c,d) $R=6$.}
\end{figure}

For $0 \le R \lessapprox 4.90$ the dynamics has the characteristic
property that the state space $\Gamma$ is divided in separate
intervals $[ ( m - 1 )\pi , ( m + 1 ) \pi ]$ with $m=0,1,2,\ldots$. In
each of these intervals occurs the dynamical scenario which has been
described so far. At $R \approx 4.90$ it happens for the first time
that previously separated intervals are linked together, so that a new
dynamical behavior becomes possible. For $4.90 \lessapprox R
\lessapprox 5.30$ the Figs.~\ref{cb}a and \ref{cb}b show that there
exist, for instance, limit cycles of period 2 in different intervals
although the constant initial function $q(t)= -2$ (dashed--dotted) $-1$
(dashed), $1$ (dotted), and $2$ (solid) was chosen in the interval $[-
\pi, + \pi]$. Thus the transient dynamics occurs in different
intervals, whereas the asymptotic dynamics is restricted to one of
these intervals.\\

\begin{figure}[ht]
\centerline{\epsfig{figure=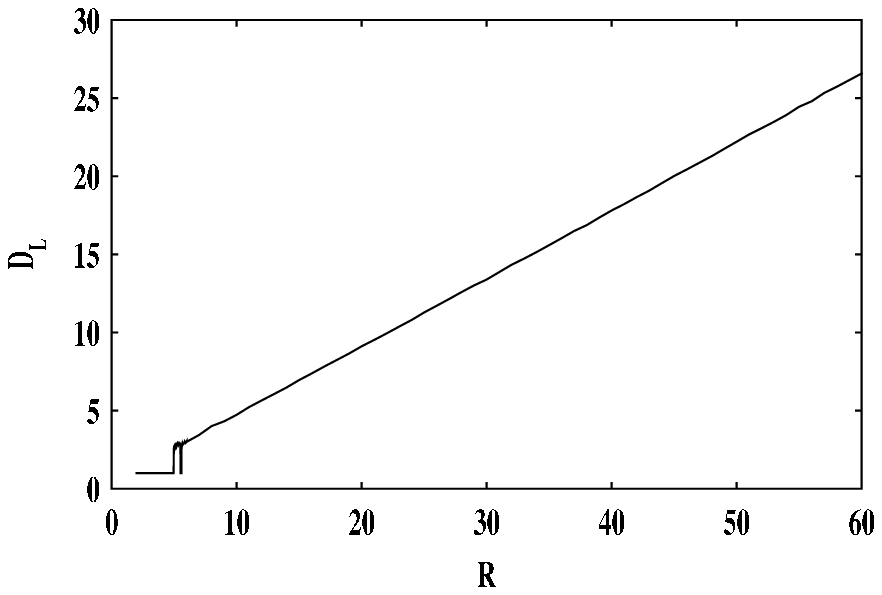,width=10cm,angle=0}}
\caption{\label{chaoatt} Lyapunov dimensions of the chaotic attractors
increasing linearly with the effective control parameter $R$.}
\end{figure}

At $R \approx 5.30$ occurs another instability to chaotic
behavior. However, now the chaotic dynamics is no longer restricted to
one of the above mentioned intervals, but it relates the previously
separated intervals. For a certain time the system dynamics remains
restricted to one of these intervals and moves then to the next
interval (see Figs.~\ref{cb}c and \ref{cb}d). Thereby the time
duration of the system within one interval differs from interval to
interval. Such a dynamics is called {\it phase slipping} or {\it cycle
slipping} \cite{Wolfgang}.  All numerical investigations above $R
\approx 5.30$ show that only the phase slipping behavior is stable.
Analyzing the Lyapunov spectrum $\lambda_1 \ge \lambda_2 \ge \ldots$,
the Lyapunov dimension
\begin{equation}
D_L = j + {\displaystyle \frac{\displaystyle
\sum_{i=1}^{j}\lambda_i}{| \lambda_j |}} \, ,
\hspace*{1cm} \sum_{i = 1}^j \lambda_i \ge 0 \, , \hspace*{1cm}
\sum_{i=1}^{j + 1} \lambda_i < 0
\end{equation}
is found to increase linearly with the control parameter $R$ (compare
Fig.~\ref{chaoatt}). Note that also other time--delayed dynamical
systems possess chaotic attractors where the envelope of the Lyapunov
dimension $D_L$ is proportional to the time delay $\tau$
\cite{Michael1,farmer,berre}. To our knowledge a theoretical
explanation for this universal phenomenon, which could predict the
system specific slopes from the respective delay differential
equations, is still lacking.

\section{Conclusion}

Here we have demonstrated by the example of the phase--locked loop with
time delay that an adequate combination of different analytical and
numerical investigation methods reveals different aspects of the rich
dynamical behavior in time--delayed nonlinear systems. The multiple
scaling method allows to derive the normal form for the Hopf
bifurcation. Phase portraits resulting from different initial
functions are capable of detecting the splitting of a limit cycle
indicating thereby symmetry--breaking bifurcations. A period--doubling
is qualitatively indicated in the power spectrum, whereas the
Lyapunov spectrum allows more quantitative statements.  In this way we
found within the numerical accuracy evidence that the period--doubling
scenario in the phase--locked loop with time delay is governed by the
Feigenbaum constant $\delta \approx 4.669$.
\section{Acknowledgment}
We cordially thank the anonymous referee for his suggestions to
improve the manuscript. In addition, we thank Hermann Haken and Arne
Wunderlin for teaching us synergetics for many years as well as
Wolfgang Wischert for introducing us to dynamical systems with time
delay a decade ago.  Furthermore, we are thankful to Elena Grigorieva
for sharing her knowledge on the multiple scaling method.  Finally,
A.P. is grateful for the hospitality of G\"unter Wunner at the
I. Institute of Theoretical Physics at the University of Stuttgart as
this article was finished there.
\end{document}